\documentclass[preprint,preprintnumbers,amsmath,amssymb]{revtex4}

\usepackage{graphicx}% Include figure files
\usepackage{dcolumn}% Align table columns on decimal point
\usepackage{bm}% bold math

\def\lsim{\lower.5ex\hbox{$\; \buildrel < \over \sim \;$}}
\def\gsim{\lower.5ex\hbox{$\; \buildrel > \over \sim \;$}}
\def\g{\ifmmode \gamma \else $\gamma$\fi}
\def\ee{\end{equation}}
\def \be{\begin{equation}}
\def \ul{\underline}

\begin{document}

\preprint{APS/123-QED}

\title{Gravitational radiation from elastic particle scattering
in models with extra dimensions}

\author{B. Koch}
 \email{koch@th.physik.uni-frankfurt.de}
\author{M. Bleicher}%
\affiliation{%
Institut f{\"u}r Theoretische Physik, J.W. Goethe
Universit{\"a}t,\\
60054 Frankfurt am Main, Germany\\
Frankfurt Institute for Advanced Studies (FIAS),\\
60054 Frankfurt am Main, Germany\\
}%

\date{\today}% It is always \today, today,

\begin{abstract}
In this paper we derive a formula for the energy loss
due to elastic N to N particle scattering in models
with extra dimensions that are compactified on a radius R. 
In contrast to a previous derivation
we also calculate additional terms that are
suppressed by factors of frequency over compactification radius.
In the limit of a large compactification radius R
those terms vanish and
the standard result for the non compactified case is recovered.
\end{abstract}

\keywords{extra dimension, gravitational radiation}
\maketitle

\tableofcontents

%%%%%%%%%%%%%%%%%%%%%%%%%%%%%%%%%%%%%%%%%%%%%%%%%%%%55

\section{Motivation}

Finding a unified theory of gravity and of the standard model
of particle physics remains an elusive
goal in quantum field theory. 
A crucial ingredient of superstringtheory
is that it needs more than 3 spatial
dimensions for their consistency.
Also
supergravity which is recognised as the low energy effective
description of an 3+d=10 dimensional M-theory 
\cite{Witten:1995ex,Horava:1995qa}.
On the one hand there are several attempts
to incorporate extra dimensions 
into
low energy field theory
\cite{Randall:1999vf,Randall:1999ee,Arkani-Hamed:1998nn,Antoniadis:1998ig,Kodama:2005cz}.
Most of these models have in common, that only
gravity is allowed to propagate into the extra dimensions.

On the other hand, classical gravitational 
waves are being looked for since a long time
and in the forthcoming years one 
finally expects to detect
this important probe for general relativity
\cite{Abramovici:1992ah}.
Therefore also classical gravitational waves within
models with extra dimensions
do provide a good framework to study
new physics.
\\

In the following chapters we derive the cross section
for gravitational radiation
in models with an even number of extra dimensions.
Although all equations are formulated for asymptotically
flat space, we keep in mind that some spatial dimensions
might be compactified. Therefore
we do not immediately drop terms that are suppressed
by higher powers of the observer distance, as
this distance is limited in some directions
by the chosen compactification radius R.

Then we show that only in the limit of large
compactification radius (or no compactification at all)
certain terms of this cross section can be
neglected, which leads to the cross section
which was already given
\cite{Cardoso:2002pa}.

%%%%%%%%%%%%%%%%%%%%%%%%%%%%%%%%%%%
%%%%%%%%%%%%%%%%%%%%%%%%%%%%%%%%%%%%55

\section{Einstein's equations with more dimensions}
\label{sec_GravWav}

Einstein's field equations with $3+d$ spatial dimensions
are a straight forward generalisation of the
3 dimensional case,
however all the indices N,M run from $0\dots 3+d$ instead
of $0\dots 3$, i.e.

\begin{equation}\label{eq_Einstein}
R_{MN}-\frac{1}{2}g_{MN}R=-8 \pi G T_{MN}.
\end{equation}

The trace of this gives
the Ricci scalar R

\begin{equation}
R(1-\frac{4+d}{2})=-8\pi G T^N_N.
\end{equation}

From this one finds the ($3+d$) dimensional
Ricci-tensor $R_{MN}$ as

\begin{equation}\label{eq_Einstein2}
R_{MN}=-8 \pi G (T_{MN}-\frac{1}{2+d} g_{MN} T^L_L),
\end{equation}

and therefore the $3+d$ dimensional gravitational source term
$S_{MN}$ is defined as

\begin{equation}\label{eq_SMN}
S_{N M} := (T_{N M}-\frac{1}{2+d} g_{MN} T^L_L).
\end{equation}

%%%%%%%%%%%%%%%%%%%%%%%%%%%%%%%%%%%%%%%%%%%%%%%%%5
\subsection{Gravitational waves in $3+d$ spatial dimensions}

Assuming small perturbations from the $3+d$ dimensional
Minkowski metric $\eta_{N M}$ with the signature $(+,-,-,-,-,...)$ we take
the following
ansatz for the metric tensor

\begin{equation}\label{eq_gAnsatz}
g_{N M}=\eta_{MN}+ h_{MN}.
\end{equation}

Inserting this ansatz into equation (\ref{eq_Einstein2}),
Einstein's field equations to first order in 
the perturbation $h$ read

\begin{equation}\label{eq_EOMforH1}
\partial_L\partial^L h_{MN}
-\partial_L\partial_N h^L_M
-\partial_L\partial_M h^L_N
+\partial_M\partial_N h^L_L= -8 \pi G S_{MN}.
\end{equation}

Here the definition of the ($3+d$) dimensional
Riemann tensor

\be\label{eq_Riemann}
\begin{array}{rl}
R_{MNOP}&=\frac{1}{2}
\left[
\partial_N \partial_P g_{MO}-
\partial_M \partial_P g_{NO}-
\partial_N \partial_O g_{MP}+
\partial_M \partial_O g_{NP}
\right]+
g_{AB}\left[
\Gamma_{OM}^A \Gamma_{NP}^B -
\Gamma_{PM}^A \Gamma_{N0}^B
\right]\\
&:=\frac{1}{2}
\left[
\partial_N \partial_P g_{MO}-
\partial_M \partial_P g_{NO}-
\partial_N \partial_O g_{MP}+
\partial_M \partial_O g_{NP}
\right]+X_{MNOP}
\end{array}
\ee

is used.
Notice that $X_{MN}$ contributes only with quadratic
or higher order terms in h.
Now we make use of the gauge invariance of Einstein's field
equations. We choose the so called
harmonic coordinate system, for which

\begin{equation}\label{eq_HarmCond1}
g^{KL}\Gamma^N_{KL}=0.
\end{equation}

Remembering the definition of the
Christoffel symbol

\be\label{eq_Christoffel}
\Gamma^A_{BC}=\frac{1}{2}
g^{AD}
\left[
\partial_C g_{BD}+\partial_B g_{CD}-\partial_D g_{BC}
\right]
\ee

and expanding (\ref{eq_HarmCond1}) to first order in h gives

\begin{equation}\label{eq_HarmCond}
\partial_L h^L_N=\frac{1}{2}\partial_L h^N_N.
\end{equation}

Using (\ref{eq_HarmCond}) in (\ref{eq_EOMforH1})
we find 

\begin{equation}\label{eq_EOMforH2}
\partial^L\partial_L h_{MN}=-16 \pi G S_{MN}.
\end{equation}

The retarded ($\tau=t-t_0-|x-y|>0$)
solution for (\ref{eq_EOMforH2}) can be found
with the $3+d$ dimensional Greens function $G^{(3+d)}(|x-y|)$

\begin{equation}\label{eq_hMN1}
h_{MN}(t,x)=N\int dt_{0} \int d^{3+d}\underline{y} 
G^{(3+d)}_{ret}(t-t_{0} ,|\underline{x}-\underline{y}|)S_{MN}(t_{0}
,\underline{y}),
\end{equation}

where N is a normalisation factor
given by 

\begin{equation}
N=-16 \pi G.
\end{equation}

The $3+d$-dimensional retarded Greens function
\cite{Hassani:1998, Galtsov:2001iv, Cardoso:2002pa} is

\be
G^{3+d}_{ret}(t,x)=-\frac{1}{(2\pi)^{4+d}}\int d^{3+d}\ul{k}
e^{i\ul{k}\ul{x}}\int dk_0 \frac{e^{-i k_0 (t-T_0)}}{k_0^2-\ul{k}^2}.
\ee

For an even number of flat extra
dimensions
this can be integrated to \cite{Galtsov:2001iv}

\be\label{eq_GFret1}
G_{ret}^{3+d}(t,x)=
\frac{1}{4 \pi} \left[
\frac{-1}{2\pi r}\frac{\partial}{\partial r}\right]^{d/2}
\left[\frac{\delta((t-t_0)-r)}{r}
\right]\,\, , d \hbox{  even}.
\ee

As the cases with an even number of extra dimensions are easier to discuss,
we will
postpone the cases with an odd number of extra dimensions.
It is convenient to bring all derivatives in (\ref{eq_GFret1}) 
to the right hand side.
Therefore we define the commutator brackets

\be
\begin{array}{lcl}\label{eq_def1}
\left[\partial_r,\frac{1}{r}\right]_{-1}&:=&1,\\
\left[\partial_r,\frac{1}{r}\right]_{0}&:=&\frac{1}{r},\\
\left[\partial_r,\frac{1}{r}\right]_{1}&:=&\frac{-1}{r^2},\\
&...&\\
\left[\partial_r,\frac{1}{r}\right]_{n}&:=&(-1)^n n! \frac{1}{r^{n+1}}.
\end{array}
\ee

Now we decompose $(\partial_r \frac{1}{r})^n \delta$
into a number ($\hbox{A}(k,n)$) times the $k^{th}$ derivative of $\delta$
with respect to its argument.

\be\label{eq_def2}
(\partial_r \frac{1}{r})^n \delta:=
\sum_{k=0}^{n} A(k,n) \delta^{(k)}
\ee

Using the definitions (\ref{eq_def1}, \ref{eq_def2}) we find a recursive formula for
(\ref{eq_GFret1}). Knowing the Greens function for $d-2$ extra dimensions
the Greens function for $d$ extra dimensions is given by

\be\label{eq_Gret3Plusd}
\begin{array}{lcl}
G_{ret}^{3+d}(t,x)&=&
\frac{1}{4\pi r}(\frac{1}{2\pi})^{d/2}\sum_{i=0}^{d/2}
\left(
\sum_{l=0}^{d/2-i}
\left|\left[ \partial_r,\frac{1}{r}\right]_{l}\right|
\hbox{A}(l+i-1,d/2-1)\frac{(l+i)!}{l! i!}
\right) \delta^{(i)}((t-t_0)-r)\\
&:=&
\frac{1}{4\pi r}(\frac{1}{2\pi})^{d/2}\sum_{i=0}^{d/2}
K(r,i) \delta^{(i)}((t-t_0)-r).
\end{array}
\ee

For the cases of $d=0,2,4,6$ the Greens functions are:
\be
\begin{array}{rl}
G_{ret}^{3}(t,x)&=\frac{\delta((t-t_0)-r)}{4\pi r},\\
G_{ret}^{3+2}(t,x)&=\frac{(\delta((t-t_0)-r)+r\delta^{(1)}((t-t_0)-r))}{8\pi^2 r^3} ,\\
G_{ret}^{3+4}(t,x)&=\frac{\delta^{(2)}((t-t_0)-r)r^2+3\delta^{(1)}((t-t_0)-r)r+3\delta((t-t_0)-r)}{16\pi^3
r^5},\\
G_{ret}^{3+6}(t,x)&=
\frac{\delta^{(3)}r^3+6\delta^{(2)}((t-t_0)-r)r^2+15\delta^{(1)}((t-t_0)-r)r+15\delta((t-t_0)-r)}
{32\pi^4 r^7}.
\end{array}
\ee

Lets assume that the observer ($\left| \ul{x}\right|$) is sitting far
away in comparison with the extension of the source ($\left| \ul{y}\right|$).
This means for $|\underline{x}|\gg|\underline{y}|$ that

\begin{equation}\label{eq_tauApprox}
\tau =t-t_0 -|\ul{x} - \ul{y}|
\approx t-t_0 - |\underline{x}|+\underline{y}
\frac{\underline{x}}{|\underline{x}|}.
\end{equation}

Keeping this in mind 
(\ref{eq_hMN1}) gives

\be\label{eq_hMN2}
\begin{array}{lcl}
h_{MN}(x)&=&
N\int dt_{0}\int d^3y_{\parallel} d^dy_{\perp} 
G_{ret}^{(3+d)}(t-t_{0},|\ul{x}-\ul{y}|)S_{MN}(t_{0},\ul{y})\\
&=&
\int dt_{0}\int d^3y_{\parallel} d^dy_{\perp} 
\frac{N}{4\pi |\ul{x}-\ul{y}|}(\frac{1}{2\pi})^{d/2}\sum_{i=0}^{d/2}
K(|\ul{x}-\ul{y}|,i) \delta^{(i)}(t-t_0-|\ul{x}-\ul{y}|)
S_{MN}(t_{0},\ul{y}).
\end{array}
\ee

Partial integration with respect to $t_0$
shuffles the derivatives from
the $\delta$ function to the source $S$

\be\label{eq_hMN3}
\begin{array}{l}
h_{MN}(x)=\\
\int dt_{0}\int d^3y_{\parallel} d^dy_{\perp} 
\frac{N}{4\pi |\ul{x}-\ul{y}|}(\frac{1}{2\pi})^{d/2}\sum_{i=0}^{d/2}
K(|\ul{x}-\ul{y}|,i) \delta(t-t_0-|\ul{x}-\ul{y}|)
(\frac{\partial}{\partial t_0})^iS_{MN}(t_{0},\ul{y}).
\end{array}
\ee

The delta function tells us at which time we
have to evaluate
$S_{MN}(t_0,\ul{y})$.
The source term $S$ is positive definite and can be expressed
by its Fourier integral

\be\label{eq_SMNfourier}
S_{MN}(\tau,\ul{y})=
\frac{1}{\sqrt{2\pi}}
\int_0^{\infty}d\omega S_{MN}(\omega,\ul{y})
e^{-i \omega \tau}+c.c.
\ee

Every derivative with respect to the time
brings down a factor $-i\omega$ from (\ref{eq_SMNfourier}).
After using (\ref{eq_tauApprox}) and integrating out
the $\delta$ functions this leads to

\be\label{eq_hMN4}
\begin{array}{lll}
h_{MN}(x)&=&
N\frac{1}{\sqrt{2\pi}4\pi}\frac{1}{(2\pi)^{d/2}}
\int d\omega 
\exp{(-i\omega(t-|\ul{x}|))}
\\
&&
\sum_{j=0}^{d/2}
K(|\ul{x}-\ul{y}|,j)
\int d^3y_{\parallel} d^dy_{\perp} 
\frac{1}{|\ul{x}|}
(i\omega)^j S_{MN}(\omega,\ul{y})\exp{(-i\omega
\ul{y}\frac{\ul{x}}{|\ul{x}|})}\\
\end{array}
\ee

The mono pole part of this gravitational
wave is found by taking $\frac{|\ul{y}|}{|\ul{x}|}\ll 1$
and therefore to lowest order 
$\frac{1}{|\ul{x}-\ul{y}|^j} \approx \frac{1}{|\ul{x}|^j}$
and  $K(|\ul{x}-\ul{y}|,j)\approx K(|\ul{x}|,j)$.

\be\label{eq_hMN0_1}
\begin{array}{lll}
h^{(0)}_{MN}(x)&=&
N\frac{1}{\sqrt{2\pi}4\pi}\frac{1}{(2\pi)^{d/2}}
\int d\omega 
\exp{(-i\omega(t-|\ul{x}|))}
\\
&&
\sum_{j=0}^{d/2}
K(|\ul{x}|,j)
\int d^3y_{\parallel} d^dy_{\perp} 
\frac{1}{|\ul{x}|} 
(i\omega)^j S_{MN}(\omega,\ul{y})\exp{(-i\omega
\ul{y}\frac{\ul{x}}{|\ul{x}|})}\\
&=&
\int d\omega 
\exp{(-i\omega(t-|\ul{x}|))} e_{MN}(\ul{x},\omega).
\end{array}
\ee

This
looks like a plane wave solution.
As the final result of this section it is shown
that the
polarisation tensor
of the induced gravitational wave
is given by

\be\label{eq_PolTen1}
\begin{array}{lrl}
e_{MN}(\ul{x},\omega)&=&
N\frac{1}{4\pi}\frac{1}{(2\pi)^{d/2}\sqrt{2\pi}}
\sum_{j=0}^{d/2}
K(|\ul{x}|,j)
\frac{1}{|\ul{x}|} 
(i\omega)^j
\int d^3y_{\parallel} d^dy_{\perp} 
S_{MN}(\omega,\ul{y})\exp{(-i\omega \ul{y}\frac{\ul{x}}{|\ul{x}|})}+c.c.\\
&=&
N\frac{1}{4\pi}\frac{1}{(2\pi)^{d/2}\sqrt{2\pi}}
\sum_{j=0}^{d/2}
K(|\ul{x}|,j)
\frac{1}{|\ul{x}|} 
(i\omega)^j
\hat{S}_{MN}(\omega)+c.c.
\end{array}
\ee

The charge conjugated part is not shown explicitly, 
to keep the formula to a readable size,
but of course it contributes to the polarisation
tensor as well.\\
In section (\ref{sec_EnMomTens})
we will explicitly calculate the source term $S_{MN}(\omega,\ul{y})$.
When doing so it is useful to remember that the "time"
coordinate corresponding to $\omega$ is $\tau$
from equation (\ref{eq_tauApprox}) and not $t$.
One would also get this result by doing two more 
Fourier transformations on $S$ and then performing
a stationary phase analysis on the exponential functions
in the integrand.

%%%%%%%%%%%%%%%%%%%%%%%%%%%%%%%%%%%%%%%%%%%
\subsection{The energy and momentum of a gravitational wave}

In this subsection 
we derive the energy momentum tensor $t_{MN}$
of the gravitational wave given in (\ref{eq_hMN0_1}).
When we derived (\ref{eq_EOMforH1}) we just took the first order
of $h$ in $R_{MN}$. Considering the approximate solution (\ref{eq_hMN1})
in the complete
field equations (\ref{eq_Einstein}) we will find
the energy momentum tensor $t_{MN}$ of the gravitational wave
(\ref{eq_hMN1}).
Expanding and rearranging (\ref{eq_Einstein}) with 
$R_{MN}^{(1)}-\frac{1}{2}\eta_{MN}R^{(1)}$
gives

\be
R_{MN}^{(1)}-\frac{1}{2}\eta_{MN}R^{(1)}=
-8\pi G\left[T_{MN}+
\frac{1}{8 \pi G}(
R_{MN}-\frac{1}{2}\eta_{MN}R-R_{MN}^{(1)}\frac{1}{2}\eta_{MN}R^{(1)})
\right].
\ee

Now we can define the energy momentum tensor of (\ref{eq_hMN2})

\be\label{eq_tMN}
t_{MN}:=
\frac{1}{8 \pi G}(
R_{MN}-\frac{1}{2}\eta_{MN}R-R_{MN}^{(1)}\frac{1}{2}\eta_{MN}R^{(1)})
\ee

and the total energy momentum tensor

\be\label{eq_tauMN}
\tau_{MN}=T_{MN}+t_{MN}
\ee

of the gravitational wave.
The total energy momentum tensor in (\ref{eq_tauMN}) 
has now two parts, the energy momentum tensor of the
source $T_{MN}$ and the energy momentum tensor $t_{MN}$
of the propagating wave itself. In order to calculate
(\ref{eq_tMN}) we need to expand 
the ($3+d$) dimensional Riemann tensor (\ref{eq_Riemann})
to $2^{nd}$ order in $h$. Therefore
we note

\be
\begin{array}{rl}
R_{AB}	&=R_{AB}^{(1)}+R_{AB}^{(2)}+\cal{O}\mit{(h^3)},\\
R 		&=g^{AB}R_{AB}\\

&=\eta^{AB}R_{AB}^{(1)}+\eta^{AB}R_{AB}^{(2)}+h^{AB}R_{AB}^{(1)}
			+\cal{O}\mit{(h^3)},\\
R_A^{A(1)}&=\eta^{AB}R_{AB}^{(1)}.
\end{array}
\ee

Using these relations, (\ref{eq_tMN}) takes the form

\be\label{eq_tMN2}
t_{MN}=\frac{1}{8 \pi G}
\left[
R_{MN}^{(2)}-\frac{1}{2}h_{MN}R^{(1)}
-\frac{1}{2}\eta_{MN}\eta^{AB}R_{AB}^{(2)}-
\frac{1}{2}\eta_{MN}h^{CD}R_{CD}^{(1)}
\right]+\cal{O}\mit{(h^3)}.
\ee

For the freely propagating gravitational wave, the metric
$g_{MN}=\eta_{MN}+h_{MN}$ satisfies the first-order
Einstein equation $R_{MN}^{(1)}=0$. The first order terms
in (\ref{eq_tMN2}) drop out and (\ref{eq_tMN2}) simplifies
to 

\be\label{eq_tMN3}
t_{MN}=\frac{1}{8 \pi G}
\left[
R_{MN}^{(2)}-\frac{1}{2}\eta_{MN}\eta^{AB}R_{AB}^{(2)}
\right]+\cal{O}\mit{(h^3)}.
\ee 

The challenge is now to derive the h dependence of
$R_{MN}^{(2)}$. First we check the $h^2$ dependence of $X_{MN}$
in (\ref{eq_Riemann}). As the Christoffel symbols ($\Gamma$) in
$X_{MN}$ (see (\ref{eq_Riemann})) contain derivatives of the metric $G_{MN}$ and 
$X_{MN}$ is proportional to $\Gamma^2$, the second order
part of $X_{MN}$ contains only terms of the form $(\partial h)(\partial h)$
in particular

\be\label{eq_NR1}
\begin{array}{rl}
X_{MN}^{(2)}=\frac{1}{4}&
\left[
(2 \partial_L h^L_S- \partial_S h^{\grave{L}}_{\grave{L}})
(\partial_M h^S_P+\partial_M h^S_P-\partial^S h_{MN})
\right.\\& \left.-
(\partial_N h_S^L+ \partial^L h_{NP}-\partial_S h_N^L)
(\partial_M h_L^S+\partial_L h_M^S-\partial^S h_{ML})
\right]
\end{array}
\ee
 
The first part of (\ref{eq_Riemann}) contributes
terms proportional to $h$ times second derivatives of $h$, in particular

\be\label{eq_NR2}
\begin{array}{rl}
\left. R_{MN}^{(2)}\right|_{\hbox{first part}}&=
h^{LS}R_{MLSN}^{(1)}\\
	&=\frac{1}{2} h^{LS}
	(\partial_M \partial_N h_{LS}- 
\partial_L \partial_M h_{NS}-
\partial_S \partial_N h_{ML}+
\partial_L \partial_S h_{MN})
\end{array}
\ee

Putting (\ref{eq_NR1}) and (\ref{eq_NR2}) together
we find the second order of $R$ in $h$

\be\label{eq_RMN2}
\begin{array}{rrl}
R_{MN}^{(2)}&=&
\left. R_{MN}^{(2)}\right|_{\hbox{first part}}+X_{MN}^{(2)}\\
&=&\frac{1}{2} h^{LS}
	(\partial_M \partial_N h_{LS}- 
\partial_L \partial_M h_{NS}-
\partial_S \partial_N h_{ML}+
\partial_L \partial_S h_{MN})+\\
&&\left[
(2 \partial_L h^L_S- \partial_S h^{\grave{L}}_{\grave{L}})
(\partial_M h^S_P+\partial_M h^S_P-\partial^S h_{MN})
\right.\\&& \left.-
(\partial_N h_S^L+ \partial^L h_{NP}-\partial_S h_N^L)
(\partial_M h_L^S+\partial_L h_M^S-\partial^S h_{ML})
\right].
\end{array}
\ee

Now we can use the plane wave solution (\ref{eq_hMN4})
and plug it respectively into (\ref{eq_RMN2}) and (\ref{eq_tMN3}).
The result will be quite lengthy and depends on some phase factors
from (\ref{eq_tMN3}). By averaging over a 
spatial region, large compared to
$1/|k|$ we can integrate out these phase factors and simplify
the result. The average is indicated by the $\langle\rangle $ brackets.
If we also remember that
$k_L k^L=0$ and that we are free to 
choose the harmonic coordinate system condition (\ref{eq_HarmCond})
we find that the trace part of (\ref{eq_tMN3}) vanishes.
Finally, we obtain the averaged energy momentum
tensor of a plane gravitational wave

\be\label{eq_tMNAv}
\begin{array}{rl}
\langle t_{MN}\rangle &=\langle R_{MN}^{(2)}\rangle \\
&=\frac{k_M k_N}{16 \pi G}(\langle e^{SL*}(\ul{x},\tau)e_{SL}(\ul{x},\tau)\rangle -
\frac{1}{2}|\langle e^L_L\rangle |^2),
\end{array}
\ee

in dependence of the polarisation tensor
$e_{MN}$.
This polarisation tensor again depends on the
energy momentum tensor of a given source.
Such a 
source tensor for elastic collisions is derived
in the next section.

%%%%%%%%%%%%%%%%%%%%%%%%%%%%%%%%%%%%%%%%%%%%%%%%%%%%%%%%%%%%%%%%%%%%%%%%%%%%%
\section{Energy momentum tensor of an elastic collision}
\label{sec_EnMomTens}

In this section 
we focus on
the energy momentum tensor
of colliding standard model particles. This tensor
is needed, because it enters the source term 
for the gravitational wave (\ref{eq_SMNfourier}).
In the ADD model \cite{Arkani-Hamed:1998rs} all standard model
particles are confined to the brane and the total energy momentum
tensor for one of these particles can be defined using 
a delta function \cite{Giudice:1998ck,Han:1998sg}

\be
T_{MN}(x)=\eta^{\mu}_M \eta^{\nu}_N
T_{\mu \nu}(x)\delta^d(x_{\perp}).
\ee

In other models like those with universal extra dimensions
\cite{Antoniadis:1990ew,Lykken:1996fj} this delta
function restriction is not needed, but the discussion
made in this section is easily translated to models of this type 
as well.
One can decompose the energy momentum tensor
into an incoming and outgoing part

\be
T_{MN}=T_{MN}^{(in)}+T_{MN}^{(out)}.
\ee

The incoming and outgoing energy momentum tensors are given in terms
of the 4 momenta of $C$ colliding particles

\be
\begin{array}{rl}
T_{MN}^{(in)}&=
\delta^{(d)}(x_{\perp})\eta_{M\mu}\eta_{N\nu}\sum\limits_{j=1}^{C}
\frac{P_{(j)}^{\mu}P_{(j)}^{\nu}}{P_{(j)}^{0}}
\delta^{(3)}(x_{\parallel}-v_{(j)\parallel}t) \theta(-t)\\
&=:\delta^{(d)}(x_{\perp})\eta_{M\mu}\eta_{N\nu}T_{(in)}^{\mu\nu}\\
T_{MN}^{(out)}&=
\delta^{(d)}(x_{\perp})\eta_{M\mu}\eta_{N\nu}\sum\limits_{j=1}^{C}
\frac{P_{(j)}^{\mu}P_{(j)}^{\nu}}{P_{(j)}^{0}}
\delta^{(3)}(x_{\parallel}-v_{(j)\parallel}t) \theta(t)\\
&=:\delta^{(d)}(x_{\perp})\eta_{M\mu}\eta_{N\nu}T_{(out)}^{\mu\nu}.
\end{array}
\ee

The source term (\ref{eq_SMN}) for incoming states
gives 

\be\label{eq_SMNin}
\begin{array}{rl}
S_{MN}^{(in)}(t,x)&=T_{MN}^{(in)}-\frac{1}{2+d}\eta_{MN}T^{(in)L}_L\\
&=\delta^{(d)}(x_{\perp})
(\eta_{M\mu}\eta_{N\nu}-\frac{1}{2+d}\eta_{MN}\eta_{\mu\nu})
T^{\mu\nu}_{(in)}(t,x_{\parallel}).
\end{array}
\ee

The incoming and the outgoing $S_{MN}$ will now be the
used as a source term for the induced
 gravitational wave (\ref{eq_hMN0_1}).
In order to know the polarisation tensor
of the wave (\ref{eq_hMN0_1})
we have to perform this $3+d$ dimensional $y$ integral

\be\label{eq_SMNomega}
\begin{array}{lrl}
\hat{S}_{MN}(\omega)&=&
\int d^3y_{\parallel} d^dy_{\perp}
S_{MN}(\omega,y)\exp {(-i\omega \ul{y}\frac{\ul{x}}{|\ul{x}|})}\\
&=&
\frac{1}{\sqrt{2\pi}}\int d\tilde{t}\int d^3y_{\parallel} d^dy_{\perp}
S_{MN}(\tilde{t},y)\exp {(-i\omega
(\tilde{t}+\ul{y}\frac{\ul{x}}{|\ul{x}|}))}.\\
\end{array}
\ee

For the incoming particles (as well as for the outgoing particles) the 
delta function in (\ref{eq_SMNin}) helps us to do this 
integral and the last part
of (\ref{eq_SMNomega}) reads

\be\label{eq_intSMN}
\begin{array}{rl}
\int d^3y_{\parallel} d^dy_{\perp}
&S_{MN}(\tilde{t},y)\exp {(-i\omega \ul{y}\frac{\ul{x}}{|\ul{x}|})}
=\,
\int d^3y_{\parallel} d^dy_{\perp} S_{MN}^{(in)}(\tau,\ul{y})
\exp {(-i\omega \ul{y}\frac{\ul{x}}{|\ul{x}|})}\\
=\,& (\eta_{M\mu}\eta_{N\nu}-\frac{1}{2+d}\eta_{MN}\eta_{\mu\nu})
\sum\limits_{j=1}^{C}
\frac{P_{(j)}^{\mu}P_{(j)}^{\nu}}{P_{(j)}^{0}}\int d^3y_{\parallel} 
\delta^{(3)}((y_{\parallel}-v_{(j)\parallel \tau})\theta(-\tau))
\exp {(-i\omega \ul{y}\frac{\ul{x}}{|\ul{x}|})}\\
=:& (\eta_{M\mu}\eta_{N\nu}-\frac{1}{2+d}\eta_{MN}\eta_{\mu\nu})
\sum\limits_{j=1}^{C}
\frac{P_{(j)}^{\mu}P_{(j)}^{\nu}}{P_{(j)}^{0}}J^{(in)}.
\end{array}
\ee

After some transformations,
$J^{(in)}$ 
can be brought to 
a form compatible with the Fourier decomposition
of $h_{MN}$:

\be\label{eq_J1}
\begin{array}{rl}
J^{(in)}&=\int d^3y_{\parallel} 
\delta^{(3)}((y_{\parallel}-v_{(j)\parallel} \tau)\theta(-\tau))
\exp {(-i\omega \ul{y}\frac{\ul{x}}{|\ul{x}|})}\\
&=\int d^3y_{\parallel}\int \frac{d^3k_{\parallel}}{(2\pi)^3}
\exp{(ik_{\parallel}(y_{\parallel}-v_{(j)}\tau))}\int \frac{d\omega_0}{-2\pi
i}
\frac{e^{-i\omega_0 \tau}e^{-i\omega \ul{y}\frac{\ul{x}}{|\ul{x}|}}}
{\omega_0-i\epsilon}\\
&=\int d^3y_{\parallel}\int \frac{d^3k_{\parallel}}{(2\pi)^3}
\exp{(ik_{\parallel}(y_{\parallel}))}\int \frac{d\omega_0}{-2\pi i}
\frac{e^{-i(\omega_0 +v_{(j)}k_{\parallel})\tau}e^{-i\omega
\ul{y}\frac{\ul{x}}{|\ul{x}|}}}
{\omega_0-i\epsilon}\\
&=\int d^3y_{\parallel}\int \frac{d^3k_{\parallel}}{(2\pi)^3}
\exp{(ik_{\parallel}(y_{\parallel}))}\int \frac{d\tilde{\omega}}{-2\pi i}
\frac{e^{-i(\tilde{\omega})\tau}e^{-i\omega \ul{y}\frac{\ul{x}}{|\ul{x}|}}}
{\tilde{\omega}-k_{\parallel}v_{(j)}-i\epsilon}\\
&=\int d\tilde{\omega}
e^{-i\tilde{\omega}\tau}
\int \frac{d^3k_{\parallel}}{-i (2\pi)^4}\int d^3y_{\parallel}
\frac{e^{ik_{\parallel}y_{\parallel}}e^{-i\omega
\ul{y}\frac{\ul{x}}{|\ul{x}|}}}
{\tilde{\omega}-k_{\parallel}v_{(j)}-i\epsilon}.
\end{array}
\ee

Here, first the Fourier transform of the $\delta$ and
$\theta$ function is used, then the terms under the integrals
are rearranged
and the substitution $\tilde{\omega}:=\omega_0+k_{\parallel}v_{(j)}$
is made.
Now the Fourier
definition of the $\delta$ function is used in order to
get rid of the two three dimensional integrals

\be\label{eq_Jin}
\begin{array}{rl}
J^{(in)}&=
\int d\tilde{\omega} 
e^{-i\tilde{\omega}\tau}
\int \frac{d^3k_{\parallel}}{-i (2\pi)^4}
\frac{1}{\tilde{\omega}-k_{\parallel}v_{(j)}-i\epsilon}
\int d^3y_{\parallel}
\exp{(-i(\omega\frac{x_{\parallel}}{|x_{\parallel}|}-k_{\parallel})y_{\parallel})}\\
&=
\int d\tilde{\omega}
e^{-i\tilde{\omega}\tau}
\frac{1}{-i2 \pi}
\frac{1}{\tilde{\omega}-\ul{k}v_{(j)}-i \epsilon}.\\
\end{array}
\ee

From 
$\underline{k}v_{(j)}=k_{\parallel}v_{(j)}$
we see that $k_{\parallel}$ can be replaced by
$\underline{k}$.
For outgoing particles the procedure is the same, one
just has to use the Fourier transform of  $\theta(-t)$

\be\label{eq_Jout}
J^{(out)}= -
\int d\tilde{\omega}
e^{-i\tilde{\omega}\tau}
\frac{1}{-i2 \pi}
\frac{1}{\tilde{\omega}- \underline{k}v_{(j)}+i \epsilon}.
\ee

We see that the difference between the incoming and outgoing $J$ can
be expressed by a change of the sign of $J$ and $\epsilon$.
These results can be plugged back into (\ref{eq_intSMN}).
For high energetic particles the denominator is
$P^0_{(j)}(\omega-\ul{k}v_{(j)})=k\cdot P_{(j)}$.
As this is $>0$ we can drop the $\epsilon$.
Using (\ref{eq_SMNomega}, \ref{eq_intSMN}, \ref{eq_Jin}, \ref{eq_Jout})
one sees that
the source terms are given by:

\be\label{eq_intSMN2}
\hat{S}^{(in)}_{MN}(\omega)=:
(\hat{T}^{(in)}_{MN}-\eta_{MN}\hat{T}^{(in)L}_{L})=
(\eta_{M\mu}\eta_{N\nu}-\frac{1}{2+d}\eta_{MN}\eta_{\mu\nu})
\sum\limits_{j=1}^{C}
\frac{P_{(j)}^{\mu}P_{(j)}^{\nu}}{P_{(j)}k}
\ee

For the outgoing particles this reads

\be\label{eq_intSMN3}
\hat{S}^{(out)}_{MN}(\omega)=:
(\hat{T}^{(out)}_{MN}-\eta_{MN}\hat{T}^{(out)L}_{L})=
-(\eta_{M\mu}\eta_{N\nu}-\frac{1}{2+d}\eta_{MN}\eta_{\mu\nu})
\sum\limits_{j=1}^{C}
\frac{P_{(j)}^{\mu}P_{(j)}^{\nu}}{P_{(j)}k}.
\ee

%%%%%%%%%%%%%%%%%%%%%%%%%%%%%%%%%%%%%%%%%%%%%%%%%%%%%%%%%%%%
\section{Gravitational radiation from elastic scattering}
\label{sec_GravRad}

Based on the discussion
in (\ref{sec_GravWav}) and (\ref{sec_EnMomTens}) we will now
 calculate the classically radiated
energy into gravitational waves from an elastic scattering.

%%%%%%%%%%%%%%%%%%%%%%%%%%%%%%%%%%%%%%%%%%%%%%%%%%%%
\subsection{Radiated energy and the energy momentum tensor}
\label{subsec_RadEnMomTens}

The momentum $P^i$ of an extended object is defined as the volume
integral over the density of the $t^{0i}$ component of the energy
momentum tensor. In $3+d$ dimensions this is

\be\label{eq_P^i}
P^i=\int_V d^{3+d}x t^{0i}.
\ee

The energy change in time $\frac{dE}{d\tau}$ of a system
can be rewritten by using the conservation
of the energy momentum tensor

\be\label{eq_dEdtau}
\frac{dE}{d\tau}=\int_V d^{3+d}x\,\partial_0 t^{00}
=\int_V d^{3+d}x\, \partial_i t^{0i}
=\partial_i P^i.
\ee

Applying Gauss law to $\partial_{i}P^i$ and using
(\ref{eq_P^i}) gives

\be\label{eq_diPi}
\partial_i P^i=\int_V d^{3+d}x \, \partial_i P^i=
\int_{\cal{O}\mit{(V)}}dS \,n_i t^{0i}=
\int_{\cal{O}\mit{(V_E)}}d\Omega \, |x|^{2+d}n_i t^{0i}.
\ee

By differentiating (\ref{eq_dEdtau}) after $d\Omega$,
averaging over the space
and integrating over $d\tau$ we get from
(\ref{eq_diPi}) the 
average energy radiated into the space-segment $d\Omega$

\be\label{eq_dEdOm}
\frac{d\langle E\rangle }{d\Omega}
=\int d\tau \frac{\langle \partial_i P^i\rangle }{d\Omega}
=\int d\tau |x|^{2+d} n_i \langle t^{0i}\rangle .
\ee

%%%%%%%%%%%%%%%%%%%%%%%%%%%%%%%%%%%%%%%%%%%%%%%%%%%%%%%%%
\subsection{Radiated gravitational energy}
\label{subsec_RadGravEn}

Using the  general
relation between radiated energy and the energy momentum 
tensor $t^{MN}$ (see (\ref{subsec_RadEnMomTens}))
we will now quantify how much energy
is radiated away by a gravitational wave. Therefore one has to
plug the energy momentum tensor of this wave (\ref{eq_tMNAv}) into 
equation (\ref{eq_dEdOm}). 
In the Fourier formulation of (\ref{eq_dEdOm}) we 
use (\ref{eq_tMNAv}) and $k_0^2=k_i^2=\omega$

\be
\begin{array}{lrl}
\frac{dE}{d\Omega}&=&
\frac{1}{2\pi}\int \int \int d\tau d\tilde{\omega} d\omega
|\ul{x}|^{2+d} \frac{\tilde{\omega}\omega}{16 \pi G}
(\langle e^{SL*}(\ul{x},\omega)e_{SL}(\ul{x},\tilde{\omega})\rangle -
\frac{1}{2}\langle e^{L*}_L(\ul{x},\omega)\rangle \langle e^{L}_L(\ul{x},\tilde{\omega}\rangle )
e^{i\tau(\tilde{\omega}-\omega)}\\
&=&
\int d\omega
|\ul{x}|^{2+d} \frac{\omega^2}{16 \pi G}
(\langle e^{SL*}(\ul{x},\omega)e_{SL}(\ul{x},\omega)\rangle -
\frac{1}{2}|\langle e^L_L(\ul{x},\omega)\rangle |^2).
\end{array}
\ee

Now we can bring the $d\omega$ to the left side and get

\be\label{eq_dEdOmdom}
\begin{array}{rl}
\frac{dE}{d\Omega d\omega}&=|x|^{2+d} n_i \langle t^{0i}\rangle \\
&=|x|^{2+d} \frac{\omega^2}{16
\pi}(\langle e^{SL*}(\ul{x},\omega)e_{SL}(\ul{x},\omega)\rangle -
\frac{1}{2}|\langle e^L_L\rangle |^2).
\end{array}
\ee

We use the relation $\omega=|k^0|=|n_i k^i|$.
From (\ref{eq_PolTen1} we get
the polarisation tensors $e_{MN}$ of the 
radiated gravitational wave, 

\be\label{eq_PolTen2}
\begin{array}{ll}
\langle e_{MN}(\ul{x},\omega)\rangle &=\,
N \frac{1}{4 \pi} \frac{1}{(2\pi)^{d/2}\sqrt{2\pi}}
\hat{S}_{MN}(\omega)
\langle \sum_{j=0}^{d/2}K(|\ul{x}|,j)\frac{1}{|\ul{x}|}(i\omega)^j\rangle .
\end{array}
\ee

Here we define
$\hat{S}_{MN}(\omega):=
(\hat{T}_{MN}(\omega)-\frac{1}{2+d}\eta_{MN}\hat{T}^L_L(\omega))$,
which is the Fourier transform
of the $(\hat{S}_{MN}(\tau)^{(in)}+\hat{S}_{MN}(\tau)^{(out)})$ we know
from equation (\ref{eq_intSMN}).
Let us first calculate the $e^{MN} e_{MN}^*$ part of (\ref{eq_dEdOmdom})
by using (\ref{eq_PolTen2})

\be\label{eq_eMNeMN}
\begin{array}{rl}
e^{SL*}(\ul{x},\omega)e_{SL}(\ul{x},\omega)=&
\frac{N^2}{32 \pi(2\pi)^{d}}
\sum\limits_{j,k=0}^{d/2}\langle K(|\ul{x}|,j)K(|\ul{x}|,k)
\frac{1}{|\ul{x}|^2}(i\omega)^{j+k}\rangle 
\,\hat{S}^{SL}(\omega)\hat{S}^*_{SL}(\omega)\\
=&
\frac{8 G^2}{\pi (2\pi)^d} 
\sum\limits_{j,k=0}^{d/2}\langle K(|\ul{x}|,j)K(|\ul{x}|,k)
\frac{1}{|\ul{x}|^2}(i\omega)^{j+k}\rangle \\&\,\,\,\,
\,\,\,\,\,\,\,\,\,\,\,\,\,\,\,\,\,\,\,\,\,\,\,\,
\,\,\,\,\,\,\,\,\,\,\,\,\,\,\,\,\,\,
(\hat{T}^{SL}(\omega)\hat{T}^*_{SL}(\omega)
-\frac{d}{(2+d)^2}|T^K_K|^2).
\end{array}
\ee

Proceeding the same way with $|e^L_L|^2$ we find

\be\label{eq_eLL2}
\begin{array}{rl}
|e^L_L|^2
=&
\frac{8 G^2}{\pi (2\pi)^d} 
\sum\limits_{j,k=0}^{d/2}\langle K(|\ul{x}|,j)K(|\ul{x}|,k)
\frac{1}{|\ul{x}|^2}(i\omega)^{j+k}\rangle \,\, 
\hat{S}^N_N \hat{S}^{L*}_L\\
=&
\frac{8 G^2}{\pi (2\pi)^d} 
\sum\limits_{j,k=0}^{d/2}\langle K(|\ul{x}|,j)K(|\ul{x}|,k)
\frac{1}{|\ul{x}|^2}(i\omega)^{j+k}\rangle \,\, 
|T^L_L|^2(\frac{2}{2+d})^2.
\end{array}
\ee

Evaluating the $T$ terms in (\ref{eq_eLL2},\,\,\ref{eq_eMNeMN})
separately leads to

\be
\hat{T}^{SL}\hat{T}^*_{SL}=
(\hat{T}^{(in)SL}+\hat{T}^{(out)SL})
(\hat{T}^{(in)*}_{SL}+\hat{T}^{(out)*}_{SL}).
\ee 

In the notation of (\ref{eq_intSMN2}, \ref{eq_intSMN3}) this will be rather
lengthy. But we can take Sums ($\sum_I..$) over all involved states 
instead of initial and final states separately ($\sum_i..+\sum_j..$) and
use that every outgoing state brings one $-$ sign. After defining

\be
\eta_I=\left\{
\begin{array}{l}
+1\,\, I\,\hbox{ in initial state}\\
-1\,\, I\,\hbox{ in final state},
\end{array}
\right.
\ee

we have 

\be
\hat{T}^{MN}=(\eta^{M\mu}\eta^{N\nu}
\sum_I\frac{P_{(I)\mu}P_{(I)\nu}\eta_I}{k P_{(I)}}).
\ee

In this notation we find that

\be
\hat{T}^{SL}\hat{T}^*_{SL}=
\sum_{I,J}\frac{(P_{(I)}^{\mu}P_{(J)\mu})^2\eta_I\eta_J}{(P_{(I)}k)(P_{(J)}k)},
\ee

and that

\be
\hat{T}_L^{L}\hat{T}^{S*}_{S}=
\sum_{I,J}\frac{P_{(I)}^2\, P_{(J)}^2\eta_I\eta_J}{(P_{(I)}k)(P_{(J)}k)}.
\ee

The last two equations can be put into (\ref{eq_eMNeMN} and
\ref{eq_dEdOmdom})
to derive the energy carried away by induced gravitational
radiation

\be\label{eq_dEdOmdom3}
\begin{array}{rl}
\frac{dE}{d\Omega d\omega}
=&
\frac{G\left|\ul{x}^{2+d}\right|}{2\pi^2 (2\pi)^d} 
\sum\limits_{j,k=0}^{d/2}\langle K(|\ul{x}|,j)K(|\ul{x}|,k)
\frac{1}{|\ul{x}|^2}(i\omega)^{j+k}\rangle \\&\,\,\,\,\,\,
\,\,\,\,\,\,\,\,\,\,\,\,\,\,\,\,\,\,\,\,\,\,\,\,\,\,\,\,\,\,
\,\,\,\,\,\,\,\,\,\,\,\, 
\sum\limits_{I,J}\frac{\eta_I\eta_J}{(P_{(I)}k)(P_{(J)}k)}
\left[
(P_{(I)}^{\mu}P_{(J)\mu})^2-\frac{1}{2+d}P_{(I)}^2\, P_{(J)}^2
\right]\\
\end{array}.
\ee

In the second step all the
simplifying definitions (\ref{eq_eLL2},
\ref{eq_eMNeMN}, \ref{eq_Gret3Plusd})
are used .
The ($3+d$) dimensional gravitational
constant G has a (d) dependent mass dimension.
This becomes more obvious by the definition
of the coupling G through a mass scale $M_f$

\be
G=\frac{1}{M_f^{2+d}}.
\ee

In the case of $d=0$ this gives
$G=\frac{1}{M_f^2}=\frac{1}{M_P^2}$ which is
the definition of the Planck mass.
For the cases with even extra dimensions
$d=0,2,4,6$ equation (\ref{eq_dEdOmdom3}) gives

\begin{eqnarray}
\label{eq_dEdOmdom4}
\frac{dE(d=0)}{d\Omega d\omega}
&=&
\frac{1}{M_P^2}\frac{1}{2\pi^2}\omega^2
\sum_{I,J}\frac{\eta_I\eta_J}{(P_{(I)}k)(P_{(J)}k)}
\left[
(P_{(I)}^{\mu}P_{(J)\mu})^2-\frac{1}{2}P_{(I)}^2\, P_{(J)}^2
\right]\nonumber\\
\frac{dE(d=2)}{d\Omega d\omega}
&=&
\frac{1}{M_f^4}\frac{1}{8\pi^4}
\left(\omega^4+2\frac{\omega^3}{|\ul{x}|}+\frac{\omega^2}{|\ul{x}|^2}\right)
\sum_{I,J}\frac{\eta_I\eta_J}{(P_{(I)}k)(P_{(J)}k)}
\left[
(P_{(I)}^{\mu}P_{(J)\mu})^2-\frac{1}{4}P_{(I)}^2\, P_{(J)}^2
\right]\nonumber\\
\frac{dE(d=4)}{d\Omega d\omega}
&=&
\frac{1}{M_f^6}\frac{1}{32\pi^6}
\left(\omega^6+6\frac{\omega^5}{|\ul{x}|}+15\frac{\omega^4}{|\ul{x}|^2}
+18\frac{\omega^3}{|\ul{x}|^3}+9\frac{\omega^2}{|\ul{x}|^4}\right)\\
&&\,\,\,\,\,
\sum_{I,J}\frac{\eta_I\eta_J}{(P_{(I)}k)(P_{(J)}k)}
\left[
(P_{(I)}^{\mu}P_{(J)\mu})^2-\frac{1}{6}P_{(I)}^2\, P_{(J)}^2
\right]\nonumber\\
\frac{dE(d=6)}{d\Omega d\omega}
&=&
\frac{1}{M_f^8}\frac{1}{128\pi^8}
\left(\omega^8+12\frac{\omega^7}{|\ul{x}|}+66\frac{\omega^6}{|\ul{x}|^2}
+210\frac{\omega^5}{|\ul{x}|^3}+405\frac{\omega^4}{|\ul{x}|^4}
+450\frac{\omega^3}{|\ul{x}|^5}+225\frac{\omega^2}{|\ul{x}|^6}
\right)\nonumber\\
&&\,\,\,\,\,
\sum_{I,J}\frac{\eta_I\eta_J\left[
(P_{(I)}^{\mu}P_{(J)\mu})^2-\frac{1}{8}P_{(I)}^2\, P_{(J)}^2
\right]
}
{(P_{(I)}k)(P_{(J)}k)}\nonumber.
\end{eqnarray}

%%%%%%%%%%%%%%%%%%%%%%%%%%%%%%%%%%%%%%%%%%%%%%%%%%%%%%%%%
\subsection{Interpretation and physical relevance
of the obtained cross sections}
\label{subsec_Discussion}

In the limit of no extra dimensions, (\ref{eq_dEdOmdom4}) agrees 
with \cite{Weinberg:1972}.
For $d\neq 0$ there are several terms contributing:
There is always one with a $\omega^{d+2}$ dependence and 
there are terms with the same mass-dimension, but containing
a conspicuous looking $|\ul{x}|$ dependence
$\frac{\omega^{d+2-i}}{|\ul{x}|^i}$. For a
uncompactified $4+d$ dimensional space-time for a distant
observer those terms vanish and only the $\omega^{2+d}$
term survives. 

For compactified Large Extra Dimensions we start from
the following setup: 
The collision region for massive particles or black holes
is smaller than the compactification radius R.
For $\ul{x}< R$ equation (\ref{eq_dEdOmdom4})
holds and the $|\ul{x}|$ terms get weaker and weaker with
distance. But when the distance $|\ul{x}|$ reaches R, the attenuation 
of those terms stops
as the world starts to look again four dimensional.
So for a given frequency $\omega$ they can be replaced by
$\frac{\omega^{d+2-i}}{R^i}$. In the
ADD model \cite{Arkani-Hamed:1998rs} 
the radius is related to the Planck-mass $M_P$ and the
new fundamental mass scale $M_f$ by

\be\label{eq_ADD}
M_P^2=M_f^{2+d}R^d.
\ee

Using this relation we can estimate for
which kind of scenarios the new terms become relevant.
As the radiated energy is increasing
rapidly with $\omega$ some cut off has to be used
to estimate the amount of gravitationally
radiated energy. In a 2 to 2 particle process
emitting gravitational radiation this cut is at least reached
as soon as the gravitational radiation takes away the invariant 
energy  $\sqrt{s}/2$ of one of the participants.
Strongest suppression of the $1/R$ terms is reached
when we take this extreme value for $\omega$.
Limits on the compactification radius down to the
$\mu m$ range (depending on $d$) have been
derived from a large number of physical observations
\cite{Anchordoqui:2002hs,Hannestad:2001xi,Barrow:1987sr,Hossenfelder:2003jz}.
Under the condition of
\be
\omega\gg\frac{1}{R}\,\,
\hbox{or}\,\,
\sqrt{s}\gg 2 M_f \left(\frac{M_f}{M_P}\right)^{2/d},
\ee
equation (\ref{eq_dEdOmdom4}) gives the
original result from
 \cite{Cardoso:2002pa}.
This shows that the additional terms only
play a role for small $\sqrt{s}$ or
very large $M_f$. 
On the one hand for particle scattering
with invariant energy in the TeV range,
$M_f$ would have to be up to $1000$ TeV, for
the new terms to be relevant. On the
other hand the whole cross-section is suppressed
by a factor $1/M_f^{2+d}$ and would be negligible
then. Summarising one can say that
for elastic high energy N to N particle collisions
in models with large extra dimensions
the energy loss into gravitational radiation stays as
described in \cite{Cardoso:2002pa} 

\begin{eqnarray}\label{eq_dEdOmdom5}
\frac{dE}{d\Omega d\omega}
=&
\frac{1}{M_f^{2+d}}\frac{\omega^{2+d}}{2(\pi)^2(2\pi)^d}
\sum_{I,J}\frac{\eta_I\eta_J}{(P_{(I)}k)(P_{(J)}k)}
\left[
(P_{(I)}^{\mu}P_{(J)\mu})^2-\frac{1}{2+d}P_{(I)}^2\, P_{(J)}^2
\right].
\end{eqnarray}

This result is valid for elastic N$\rightarrow$N particle scattering with
high particle velocities so that the
interaction can be approximated to be instantaneous.
Equation (\ref{eq_dEdOmdom4}) was
derived from classical general relativity and gives
an quantitative idea for the gravitationally radiated
energy. A quantum calculation for example in the
ADD model was not yet performed, but is considered
to be the next step to do.

%%%%%%%%%%%%%%%%%%%%%%%%%%%%%%%%%%%%%%%%%%%%%%%%%

%%%%%%%%%%%%%%%%%%%%%%%%%%%%%%%%%%%%%%%%%%%%%%%%5
\section{Summary}
The main concern of this paper was
to derive the general energy loss formula due to 
gravitational radiation in models
with extra dimensions 
that are compactified on a radius R

\be
\begin{array}{rl}
\frac{dE(d=0)}{d\Omega d\omega}
=&
\frac{1}{M_P^2}\frac{1}{2\pi^2}\omega^2
\sum_{I,J}\frac{\eta_I\eta_J}{(P_{(I)}k)(P_{(J)}k)}
\left[
(P_{(I)}^{\mu}P_{(J)\mu})^2-\frac{1}{2}P_{(I)}^2\, P_{(J)}^2
\right]\\
\frac{dE(d=2)}{d\Omega d\omega}
=&
\frac{1}{M_f^4}\frac{1}{8\pi^4}
\left(\omega^4+2\frac{\omega^3}{|R|}+\frac{\omega^2}{|R|^2}\right)
\sum_{I,J}\frac{\eta_I\eta_J}{(P_{(I)}k)(P_{(J)}k)}
\left[
(P_{(I)}^{\mu}P_{(J)\mu})^2-\frac{1}{4}P_{(I)}^2\, P_{(J)}^2
\right]\\
\frac{dE(d=4)}{d\Omega d\omega}
=&
\frac{1}{M_f^6}\frac{1}{32\pi^6}
\left(\omega^6+6\frac{\omega^5}{|R|}+15\frac{\omega^4}{|R|^2}
+18\frac{\omega^3}{|R|^3}+9\frac{\omega^2}{|R|^4}\right)
\sum_{I,J}\frac{\eta_I\eta_J}{(P_{(I)}k)(P_{(J)}k)}
\left[
(P_{(I)}^{\mu}P_{(J)\mu})^2-\frac{1}{6}P_{(I)}^2\, P_{(J)}^2
\right]\\
\frac{dE(d=6)}{d\Omega d\omega}
=&
\frac{1}{M_f^8}\frac{1}{128\pi^8}
\left(\omega^8+12\frac{\omega^7}{|R|}+66\frac{\omega^6}{|R|^2}
+210\frac{\omega^5}{|R|^3}+405\frac{\omega^4}{|R|^4}
+450\frac{\omega^3}{|R|^5}+225\frac{\omega^2}{|R|^6}
\right)\\
&\,\,\,\,\,
\sum_{I,J}\frac{\eta_I\eta_J\left[
(P_{(I)}^{\mu}P_{(J)\mu})^2-\frac{1}{8}P_{(I)}^2\, P_{(J)}^2
\right]
}
{(P_{(I)}k)(P_{(J)}k)}.
\end{array}
\ee

Then we showed that for models with
large compactification radii (compared
to the wave length of the gravitational radiation)
this goes into 

\be
\begin{array}{rl}
\frac{dE}{d\Omega d\omega}
=&
\frac{1}{M_f^{2+d}}\frac{\omega^{2+d}}{2(\pi)^2(2\pi)^d}
\sum_{I,J}\frac{\eta_I\eta_J}{(P_{(I)}k)(P_{(J)}k)}
\left[
(P_{(I)}^{\mu}P_{(J)\mu})^2-\frac{1}{2+d}P_{(I)}^2\, P_{(J)}^2
\right],\\
\end{array}
\ee
in line with Ref.
\cite{Cardoso:2002pa}.
For small compactification radii
(and therefore large $M_f$)
the overall $\frac{1}{M_f^{2+d}}$ factor
strongly suppresses all terms.

%%%%%%%%%%%%%%%%%%%%%%%%%%%%%%%%%%%%%%%%5

\begin{acknowledgments}
The authors thank S. Hofmann, U. Harbach and
S. Hossenfelder for fruitful discussions and
the Frankfurt International Graduate School of Science (FIGSS) for financial
support through a PhD fellowship.
\end{acknowledgments}

\appendix

%%%%%%%%%%%%%%%%%%%%%%%%%%%%%%%55
\bibliography{gravrad}
\end{document}